\documentclass[aps,prc,reprint,superscriptaddress,showpacs,preprintnumbers,amssymb]{revtex4-1}
\usepackage{graphics} 
\usepackage{graphicx} 
\usepackage{epstopdf}
\usepackage{color}
\usepackage{scrtime}
\usepackage{amssymb}
\usepackage{amsmath}
\usepackage{amsfonts}

\usepackage{color,hyperref}
\definecolor{blue}{rgb}{0.0,0.0,1}
\hypersetup{colorlinks,breaklinks,
            linkcolor=blue,urlcolor=blue,
            anchorcolor=blue,citecolor=blue}

\def\2ih{$2\Im/\hbar^2$}
\def\h2i{$\hbar^2/2\Im$}
\def\beq{\begin{equation}}
\def\eeq{\end{equation}}

\def\i13{i$_{13/2}$}

\begin{document}

%\preprint{{\underline{\it REV\TeX~4-1: $^{16}$C Version 27}}}

\title{Tidal Waves in $^{102}$Pd:  A Phenomenological Analysis}


\author{A.~O.~Macchiavelli}
\affiliation{Nuclear Science Division, Lawrence Berkeley National Laboratory, Berkeley, California 94720, USA}
\author{A. D. Ayangeakaa}\altaffiliation{Present Address, Physics Division, Argonne National Laboratory, Argonne, Illinois 60439 }
\author{ S. Frauendorf}
\author{U. Garg}
\author{ M. A. Caprio}
\affiliation{Department of Physics, University of Notre Dame, Notre Dame, Indiana 46556, USA}


%\date{\thistime, \today}
\date{\today}

\begin{abstract}
Rotational and electromagnetic properties of the yrast band in $^{102}$Pd   are analyzed in terms of a phenomenological 
phonon model that includes anharmonic terms. Both the moment of inertia and $B(E2)$'s are well reproduced by the model,
providing an independent confirmation of the multi-phonon picture recently proposed.  The (empirical) dependence of the phonon-phonon 
interaction on the phonon frequency,  in Ru, Pd, and Ru isotopes,  follows the expectations from particle-vibration coupling.
\end{abstract}

%\keywords{}

\maketitle

%\section{}
%\subsection{}
%\subsubsection{}

The subject of collective vibrational motion in nuclei has been a central theme of study in nuclear  physics \cite{BM}. ÊIn particular, the existence of rather pure multi-phonon  states is still the subject of debate as the mixing with a background of quasi-particle states fragments the collective levels \cite{Garret}. In a recent work \cite{stefan}, the yrast states of spherical or weakly deformed nuclei have been described as arising from the rotation-induced condensation of aligned quadrupole phonons. Semiclassically, this condensate represents running waves on the nuclear surface (tidal waves).

Recently, $B(E2)$ reduced transition probabilities in the yrast band of Êthe nucleus $^{102}$Pd were extracted from lifetime measurements using the DSAM technique \cite{daniel}. ÊÊThis new information, together with the level energies, Êwas interpreted as evidence for tidal waves, and calculations based on the microscopic model of Ref.  \cite{stefan}  account rather well for the observed properties.  If the multi-phonon picture of the yrast band in $^{102}$Pd is indeed valid, it should  also be possible to explain its properties from the (complementary) point of view of a standard phonon model that takes into account anharmonic effects \cite{BM,brink,Das,zamfir,reagan}.  In this brief report, we present a phenomenological analysis based on such an approach.

In the simplest approximation of pure harmonic motion, the energies and the angular momenta of the yrast states 
are simply related to the number of phonons $n $ ( of energy $\hbar\omega_0$ ) by:
\begin{align}
E_{n}=n\hbar\omega _0
\end{align}
and
\begin{align}
I =2n 
\end{align}
since each quadrupole phonon carries two units of angular momentum.

Considering the effects of anharmonic terms arising from the interaction between phonons  \cite{BM,brink,zamfir,reagan}, 
Eq. (1) gets modified as follows:
\begin{align}
E_{n} \approx n\hbar\omega_0 + \frac{1}{2}V_2n(n-1) + \frac{1}{6}V_3n(n-1)(n-2)+ ...
\end{align}
which in terms of angular momentum can be written in the general form:
%\begin{align}
%E(I) \approx  ( \frac{\hbar\omega_0}{2} - \frac{V_0}{4})  I + \frac{1}{8} V_0I^2 + ... \nonumber
%\end{align}
%which can generally be written in the form
\begin{align}
E(I) \approx  a  I + bI^2 + cI^3 + ...
\end{align}
Any resemblance to a vibrational picture requires that the anharmonic coefficients should be small
compared to $a$.  
From Eq. (4), the rotational frequency, $\omega$, and the kinematic moment of inertia, $\Im^{(1)}$, can be derived as
\begin{align}
\hbar\omega \equiv \frac{\partial E(I)}{\partial I} \approx a + 2bI +3cI^2\nonumber
\end{align}
and
\begin{align}
\frac{\Im ^{(1)}}{\hbar^2} \equiv \frac{I}{\hbar\omega}  \approx  \frac{I}{a + 2bI +3bI^2}
\end{align}
respectively.

\begin{figure}
\centerline{%
\includegraphics[width=8.0cm,angle=90]{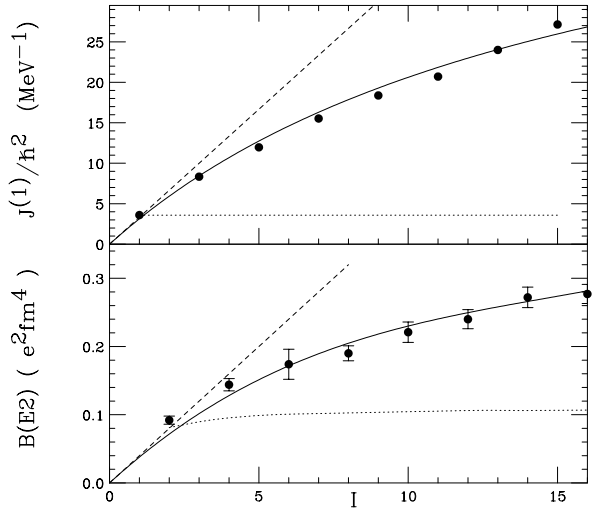}}
\caption{ Top: Moment of inertia in the yrast band in $^{102}$Pd and the results of Eq. (5). 
Bottom:  $B(E2)$ values and Eq. (6). The harmonic and rotational limits are shown in dashed and dotted lines}
\label{Fig:fig1}
\end{figure}

Along the same lines, the transition probabilities can be parametrized in a similar form \cite{brink,zamfir,reagan} to the same order

\begin{align}
B(E2) \approx \alpha I+\beta I^2+ \gamma I^3+ ...
\end{align}

In Fig. \ref{Fig:fig1}  we show the experimental data together with the fits of Eqs. (5) and (6).  
For the moment of inertia the parameters are $a = 0.254(5)$keV, $b = 0.018(1)$keV, and $c = -0.0004(4)$keV, and for the transition probabilities  $ \alpha= 4.7 (9) ~10^{-2 }e^2b^2,  \beta =-0.35(10) ~10^{-2 }e^2b^2$ , and $ \gamma =0.01(5) 10^{-2 }e^2b^2 $.  For reference, the harmonic and rotational limits are also shown in dashed and dotted lines respectively \footnote{ The reader may want to compare these results with those of Figs. 4 and 5 in Ref. \cite{daniel}}.

The good agreement with the data and the values of the coefficients indicate the applicability of the anharmonic expansion.
This  suggests the possibility of  going one step further and study the correlation between $\hbar\omega_0$ and $V_2$, obtained from the coefficients $a,b$ , and $c $ above. General arguments \footnote{See the discussion on page 338 of Ref. \cite{BM}}
indicate that as fluctuations of the spherical shape become increasingly pronounced the nucleus will eventually deform,
and in the process of this shape transition the vibrations will be subject to large anhamornicities.  Thus, we would expect that the more collective the phonon the larger the anharmonic terms, or in other words, $V_2$ to increase when $\hbar\omega_0 $ decreases.

The theory of particle-vibration coupling \cite{BM, Bes, Bertsch, Paar} provides the tools to calculate the phonon-phonon interaction $V_2$ in terms of the microscopic properties
of the basic phonons.
%as schematically illustrated in Fig. \ref{Fig:fig3}.  %
In fact,  the cases discussed in \cite{Bes,Bertsch} (based on the Lipkin model \cite{Lipkin}),  confirm our qualitative expectation and predict for example:
\begin{align}
%V_0 \approx 2\Delta- \hbar\omega_0\nonumber
V_2 \approx A- B \hbar\omega_0\nonumber
\end{align}
for the exact solution, and

\begin{align}
V_2 \approx \frac{C}{(\hbar\omega_0)^2} \nonumber
\end{align}
in the perturbative approach.

%\begin{figure}
%\centerline{
%\includegraphics[width=7.0cm, angle=0]{fig3newb.pdf}}
%\caption{  Schematic representation of the effective phonon-phonon interaction $V_2$ in Eq. (3), in terms of the microscopic structure of the phonons.  (See \cite{BM, Bes, Bertsch, Paar} for more details.)}
%\label{Fig:fig3}
%\end{figure}

\begin{figure}
\centerline{
\includegraphics[width=9.5cm, angle=0]{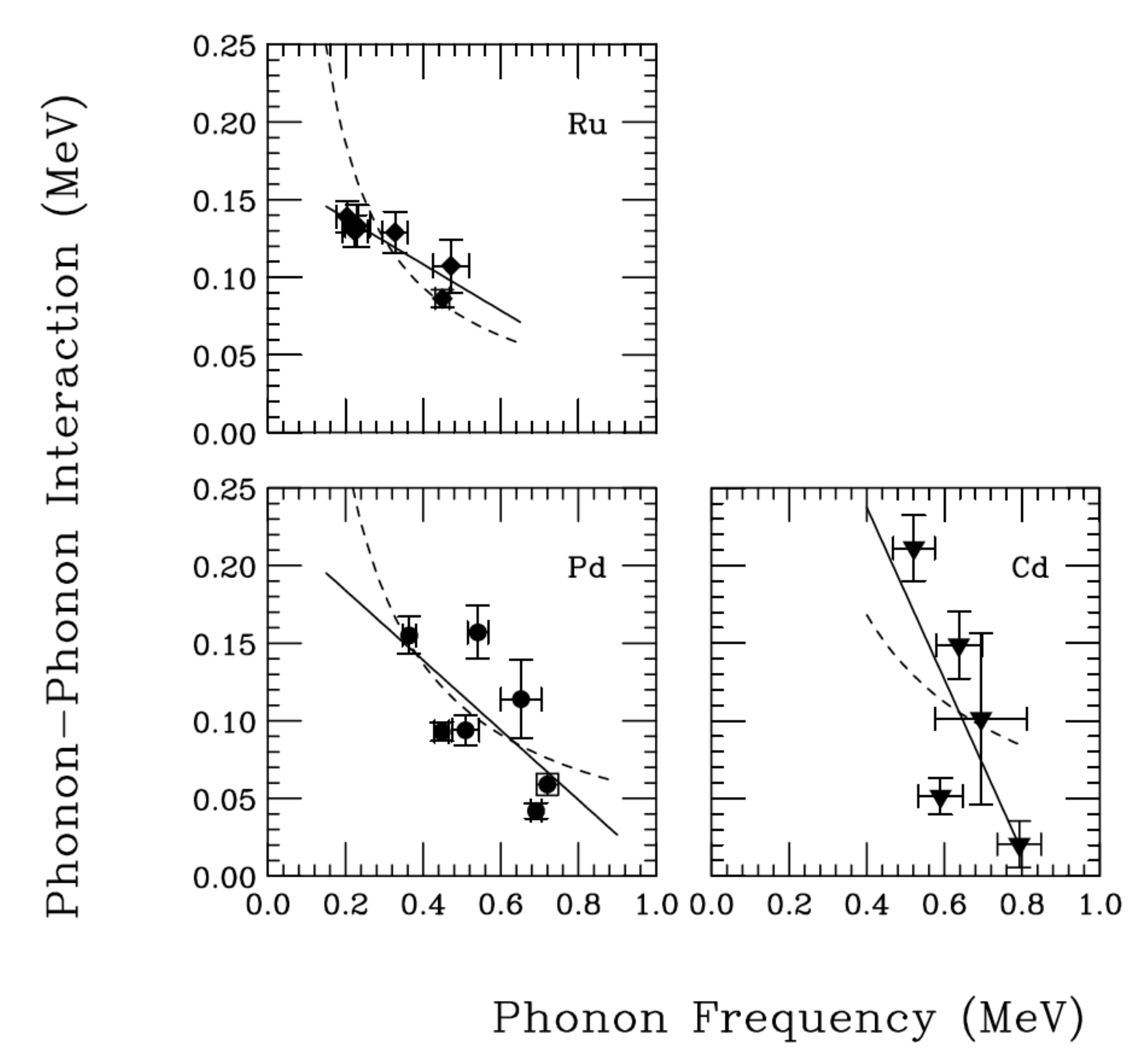}}
\caption{ Phonon-phonon interaction ($V_2$) vs. frequency ($\hbar\omega_0$) for Ru, Pd and Cd isotopes.  The lines are expectations from particle-vibration coupling models,  solid line 
Ref. \cite{Bes} and dashed line  Ref. \cite{Bertsch}.}
\label{Fig:fig3}
\end{figure}

We have extended the analysis of moments of inertia described above for $^{102}$Pd to neighboring Ru, Pd and Cd isotopes and obtained the systematics shown in Fig. \ref{Fig:fig3}.

As we have argued,  $V_2$ displays a decreasing trend with $\hbar\omega_0$ which is compared to the expressions discussed above.
While the overall behavior is reproduced, it would be of interest to see if the particle-vibration coupling approach, along the lines of the
more realistic development in \cite{Paar}, can also account for the absolute scales.     This could be the subject of a future study.

In summary,  we have analyzed the properties of the yrast band in $^{102}$Pd on the basis of a phenomenological phonon model, that includes
anharmonic terms.  Both the moment of inertia and electromagnetic properties are well reproduced by Eqs. (5) and (6), and provide a complementary confirmation of the multi-phonon picture proposed in \cite{stefan,daniel}.
The relation between $V_2$ and $\hbar\omega_0$ in Ru, Pd, and Cd nuclei appears to be accounted for by the expectations from particle-vibration coupling.

%%===========================================================================
\begin{acknowledgments}
This material is based upon work supported by the U.S. Department of Energy, Office of Science, Office of  Nuclear Physics under Contract
No.~DE-AC02-05CH11231(LBNL) and Grant No. DE-FG02-95ER40934 (UND), and by the National Science Foundation under Grant  No. PHY-1068192 (UND).
\end{acknowledgments}
%%===========================================================================

\bibliographystyle{apsrev4-1}

\bibliography{tidalphonon}

%\begin{thebibliography}{18}%

%\bibitem{BM}A. Bohr and B.R.Mottelson, Nuclear Structure Vol II, ÊChapter 6.

%\bibitem{stefan} S. Frauendorf et al.,  Int. Journal of Modern Physics E 20, 465 (2011) and arXiv-id: 0709.0254.

%\bibitem{daniel} A.D.Ayangeakaa et al. , PRL 110, 102501 (2013).

%\bibitem{brink} D.M.Brink et al., ÊPLB 19, 413 (1965).

%\bibitem{zamfir}N.V.Zamfir and R.F.Casten, PRL 75 1280 (1995).

%\bibitem{reagan}P. Reagan et al. , PRL 90, 152502 (2003).

%\bibitem{Bes} D.R.B\`es et al., Phys. Lett. B52, 253 (1974).

%\bibitem{Paar} V.Paar, Nucl. Phys. A166, 341 (1971).

%\end{thebibliography}%

\end{document}